\documentclass[%
superscriptaddress,reprint,
amsmath,amssymb,
prb,
]{revtex4-1}

\usepackage{graphicx}
\usepackage{dcolumn}
\usepackage{bm}

\usepackage{hyperref}
\usepackage{hypcap}
\hypersetup{pdftex,colorlinks=true,citecolor=red,linkcolor=blue}

\newcommand{\x}{\times}

\newcommand{\comentar}[1]{}

\usepackage{ulem} 
\usepackage[dvipsnames]{xcolor} 

\newcommand{\review}[1]{{\leavevmode\color{black}#1}}

\graphicspath{{Images/}}

\begin{document}

\title{Time-dependent forces between a swift electron and a small nanoparticle \review{within the dipole approximation}}


\author{J. Castrej\'on-Figueroa}%
\affiliation{Departamento de F\'isica, Facultad de Ciencias, Universidad Nacional Aut\'onoma de M\'exico, Ciudad Universitaria, Av. Universidad $\# 3000$, Mexico City, 04510, Mexico.}
\author{J. \'A. Castellanos-Reyes}
\affiliation{Departamento de F\'isica, Facultad de Ciencias, Universidad Nacional Aut\'onoma de M\'exico, Ciudad Universitaria, Av. Universidad $\# 3000$, Mexico City, 04510, Mexico.}
\author{C. Maciel-Escudero}
\affiliation{Departamento de F\'isica, Facultad de Ciencias, Universidad Nacional Aut\'onoma de M\'exico, Ciudad Universitaria, Av. Universidad $\# 3000$, Mexico City, 04510, Mexico.}
\affiliation{Materials Physics Center, CSIC-UPV/EHU, Donostia-San Sebasti\'{a}n, 20018, Spain.}
\author{A. Reyes-Coronado}
\affiliation{Departamento de F\'isica, Facultad de Ciencias, Universidad Nacional Aut\'onoma de M\'exico, Ciudad Universitaria, Av. Universidad $\# 3000$, Mexico City, 04510, Mexico.}
\author{R. G. Barrera}
\affiliation{Instituto de F\'isica, Universidad Nacional Aut\'onoma de M\'exico, Apartado Postal 20-364, Mexico City, 01000, Mexico.}%

\date{\today}

\begin{abstract}

\review{In this paper we calculate the time-dependent forces between a swift electron traveling at constant velocity and a metallic nanoparticle made of either aluminum or gold. We consider that the nanoparticle responds as an electric point dipole and we use classical electrodynamics to calculate the force on both the nanoparticle and the electron. The values for the velocity of the electron and the radius of the nanoparticle were chosen in accordance with electron microscopy observations, and the impact parameter was selected to fulfill the constraints imposed by the dipole approximation. We found that there are times when the force on the nanoparticle is attractive and others when it is repulsive, and show that this is due to the delayed electromagnetic response of the nanoparticle. To establish the limits of validity of our approach, we calculate the total linear momentum transfer to the nanoparticle, and compare it with results obtained, in frequency space, using the full multipole expansion of the fields induced on the nanoparticle, considering the effects of electromagnetic radiation.}



\end{abstract}

\maketitle

\section{Introduction}

The ability to manipulate objects at deep subwavelength dimensions is an interesting research area that aims to control and engineer structures at the nano and micrometer scales.\citep{Volpe,Dholakia1,VolpeBook,Romo-Herrera,Morita} Since the last century, and due to its importance in technological applications, nano and micromanipulation have been looking for techniques that allow  moving, trapping and assembling objects with accuracy. 
Optical tweezers are one example useful for the latter purpose. 
In this technique, the electromagnetic forces produced by tightly focused laser beams allow to move and hold micro objects\citep{Dholakia1,VolpeBook,Ashkin1}
, and even plasmonic nanoparticles.\citep{Oddershede,Volpe,Volpe1}
%
%
%
In this electromagnetic context, electron beams produced in transmission electron microscopes (TEM) have also shown to be potentially useful probes for guiding the motion of nanoparticles (NPs).\cite{Batson2,NanoLet,HZheng,Oleshko,TChen, YLiu, SGwo, Ticos} 
In previous works it was demonstrated that electron beams are capable to induce coalescence in separated nanoscale metal particles.\citep{Batson2,NanoLet,YLiu,TChen} 
Interestingly, the recent ability to achieve sub-angstrom resolution in the aberration-corrected scanning transmission electron microscopes (STEM),\cite{Haider,Krivanek0,Haider2,Batson,Erni} and the parallel efforts to push the spectral resolution in the range of millielectronvolts,\cite{krivanek1,krivanek2,konecna} make STEM an attractive alternative tool for nanomanipulation with high spatial and spectral resolutions. Further developments of these techniques require a full understanding of the interaction between NPs and electron beams.
%

The interaction between spherical NPs and aloof STEM electron beams has been previously addressed by calculating the total linear momentum transferred by the electron beam to the NP.\cite{GAbajo,PRB2010,PRB2016,Ultramicroscopy} These previous works were based on a classical electrodynamics approach, and the electromagnetic fields produced by the electron beam and the spherical NP were obtained by solving Maxwell's equations in the frequency space (full-retarded wave solution approach).
In addition, through a frequency-to-time Fourier transform of the electromagnetic fields, it was possible to calculate the forces upon the NP,\cite{PRB2016} showing that there are times when the force is attractive between the electron and the NP, while in others the force is repulsive. However, besides this approach is highly computational-demanding, the conclusions reached in reference\setcitestyle{numbers,square}\citep{PRB2016} regarding the nature of the repulsive forces are different to those presented here.\setcitestyle{numbers,super}

In this work we use a simple classical model to study, in time domain, the electromagnetic-interaction forces between a small metallic NP and a swift electron. We first consider the electron as a classical point particle traveling in a straight line,\cite{Note1} and model the electromagnetic response of the NP as the one of an induced electric point dipole. We then calculate the forces between the two particles (swift electron and NP) using the Lorentz-force formula. This simple description of the system allows us to understand in detail some properties of the interparticle forces and reveals their attractive or repulsive behavior at different times. Most importantly, besides the simplicity of our approach, in its range of validity \review{(large impact parameter to radius ratio)}, it quantitatively coincides with the results for the total linear momentum transfer, obtained with the full-retarded wave solution.



\section{Theoretical model: The (time-dependent) dipole approximation}\label{th}

\begin{figure}
\centering
\includegraphics[width=0.48\textwidth]{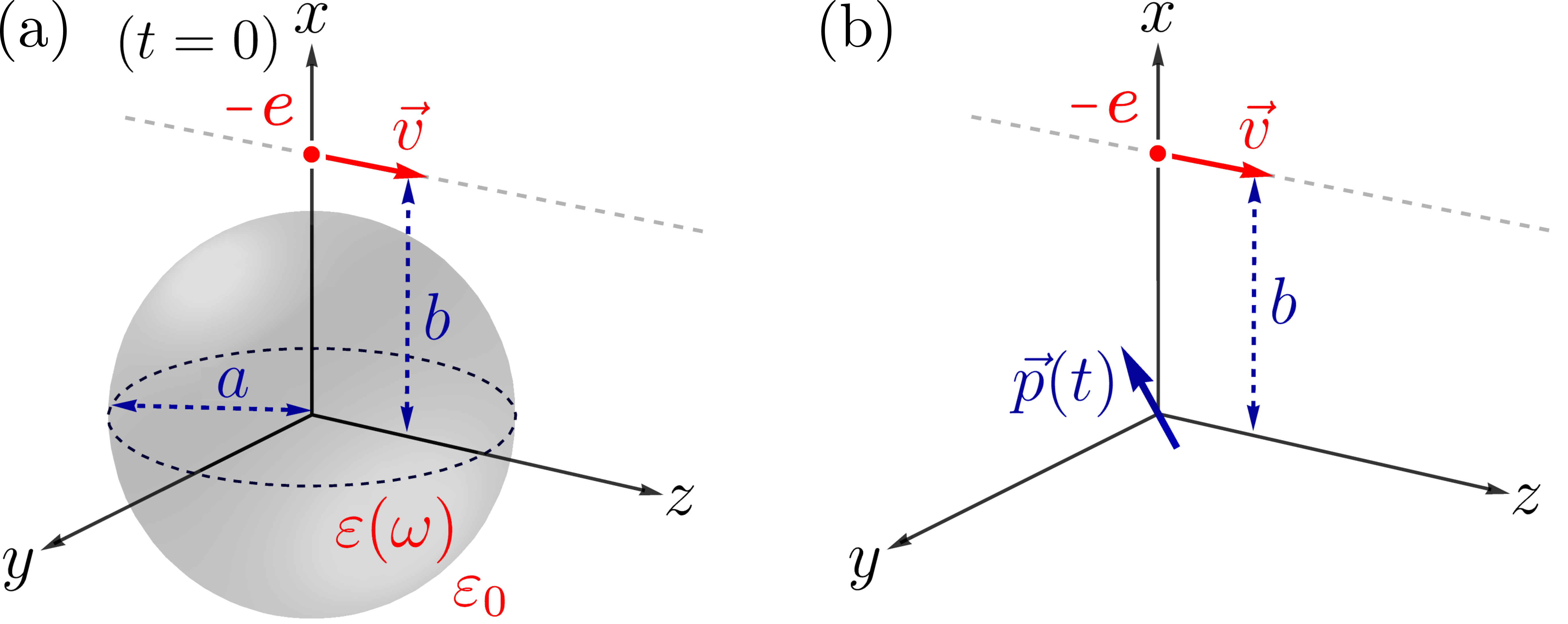}
\caption{(a) Metallic NP (grey sphere) of radius $a$ with dielectric function $\varepsilon (\omega)$, embedded in vacuum, interacting with a swift electron (red dot) traveling in the $z$-direction with constant velocity $\vec{v}$ and impact parameter $b$. (b) Schematics of the dipole approximation model: a time-dependent electric point dipole $\vec{p}(t)$ induced within the NP.}
\label{Fig1}
\end{figure}

Typical modern STEMs pump high energetic electron beams with an energy up to $400\,\text{keV}$  ($\sim0.83c$, with $c$ the speed of light). These electron beams produce low currents of the order of pA which are equivalent to a train of swift electrons traveling with constant speed, each one emitted approximately every $10^{-8}\,\text{s}$. By comparing this time interval with the typical lifetime of electronic excitations in metals ($10^{-15}\,\text{s}$),\cite{Quijada} we assume that the interaction between the electron beam and the NP can be modeled as if one single swift electron interacts with the NP.\cite{GAbajo,PRB2010,PRB2016,GAbajo99,Ferrell,Ultramicroscopy}
We also assume that the NP is a non-magnetic sphere embedded in vacuum, with radius $a$ and characterized with an isotropic homogeneous electromagnetic response $\varepsilon(\omega)$. We consider the swift electron as a point particle with electric charge $-e$, traveling with constant velocity $\vec{v}$ along the $z$ direction, and at a distance $b$ from the origin (impact parameter) which is set at the center of the NP [see Fig. \ref{Fig1}(a)]. Notice that at time $t=0$ the electron is at the position $(b,0,0)$ [see Fig. \ref{Fig1}(a)].
%

When the impact parameter is large enough compared to the radius of the NP, the NP can be regarded as an electric point dipole\cite{Ferrell} (as we will discuss in Section IV).
Within this approximation, we calculate the forces between the swift electron and the NP based on the Lorentz force. We will refer to this approach as the dipole approximation.

The induced electric dipole moment $\vec{p}(\omega)$ within the NP is determined by the following expression:
\begin{align}\label{p(w)}
\vec{p\,}(\omega) =	 \varepsilon_0 \alpha_s(\omega) \vec{E}^{\text{ext}}_0(\omega),
\end{align}
where $\varepsilon_0$ is the permittivity of vacuum, $\vec{E}^{\text{ext}}_0 (\omega)$ is the external electric field (in the frequency domain) produced by a swift electron (see Eq. (\ref{ExternalE}) in Appendix \ref{Apendix A}) evaluated at the center of the NP, and
\begin{equation}
\alpha_s(\omega)=4\pi a^3 \left[ \frac{\varepsilon(\omega) - \varepsilon_0}
{\varepsilon(\omega) + 2 \varepsilon_0} \right],
\label{staticpol}
\end{equation}
is the quasi-static polarizability.\cite{Jackson} The validity of using $\alpha_s(\omega)$ in Eq. (\ref{p(w)}) will be discussed in section IV.  

Notice that in Eq. (\ref{p(w)}) the induced electric dipole moment, $\vec{p}$, is given as a function of frequency. Through a frequency-to-time Fourier transform one obtains $\vec{p}$ as a function of time:
\begin{align}\label{dipole}
\vec{p}\,(t) =& \varepsilon_0 \int_{-\infty}^{t} \alpha_s(t-t') \vec{E}^{\text{ext}}_0(t') \text{d}t' \nonumber \\
 =& \varepsilon_0 \int_{0}^\infty \alpha_s(\tau) \vec{E}^{\text{ext}}_0(t-\tau) \text{d}\tau, 
\end{align}
where $\tau = t-t'$.
Furthermore, the force on the electric point dipole is given by:\citep{Griffiths0}
\begin{equation}\label{Fp}
\vec{F}(t)=\left[ \, \vec{p}(t) \cdot \nabla \right] \vec{E}^{\text{ext}}_0(t) + 
\dot{\vec{p}}\,(t) \times \mu_0 \vec{H}^{\text{ext}}_0(t),
\end{equation}
where $\dot{\vec{p}}$ denotes the time derivative of $\vec{p}$, $\mu_0$ is the permeability of vacuum, $\vec{E}^{\text{ext}}_0(t)$ and $\vec{H}^{\text{ext}}_0(t)$ are the external electric and H field, respectively, produced by the swift electron [Eqs. (\ref{Et}) and (\ref{Ht}) in Appendix \ref{Apendix A}] evaluated at the center of the NP. In a similar manner, the force on the swift electron is
\begin{equation}\label{Fe}
\vec{F}_\textit{e}(\vec{r}_e;t) = -e \left[ \vec{E}_p(\vec{r}_\textit{e};t) +  \vec{v} \times \mu_0 \vec{H}_p(\vec{r}_\textit{e};t) \right],
\end{equation}
where $\vec{E}_p(\vec{r}_\textit{e})$ and $\vec{H}_p(\vec{r}_\textit{e})$ are the electric and H fields, produced by the electric point dipole, evaluated at the position of the swift electron $\vec{r}_\textit{e}(t)$. We address the reader to Appendix \ref{Apendix A} for the general expressions of both, the external and the electric point dipole electromagnetic fields.


%

As pointed out in the Introduction, the interaction between a swift electron and a NP has been studied in the frequency domain by obtaining the total linear momentum, $\Delta\vec{P}$, transferred by the swift electron to the NP (full-retarded wave solution).\cite{GAbajo,PRB2010,PRB2016,Ultramicroscopy} Thus, to corroborate the validity of the dipole approximation, we calculate $\Delta\vec{P}$ for both approaches: the full-retarded wave solution and the dipole approximation. This comparison allows us to determine the accuracy of our approach.

Within the dipole approximation, the total linear momentum transferred by the swift electron to the electric point dipole is obtained by integration of Eq. (\ref{Fp}) in time:

\begin{equation}\label{DPp}
\Delta \vec{P} = \int_{-\infty}^{\infty} \vec{F} (t)\, \text{d}t.
\end{equation}
Furthermore, the total linear momentum lost by the electron is
\begin{equation}\label{DPe}
\Delta \vec{P}_e = \int_{-\infty}^{\infty} \vec{F}_e (t)\, \text{d}t,
\end{equation}
with $\vec{F}_e (t)$ given by Eq. (\ref{Fe}).

Interestingly, the dipole approximation provides valuable information and reveals interesting properties in the interaction between a swift electron and a small NP, as we discuss in the next section.

%
\begin{figure*}
\includegraphics[width=1\textwidth]{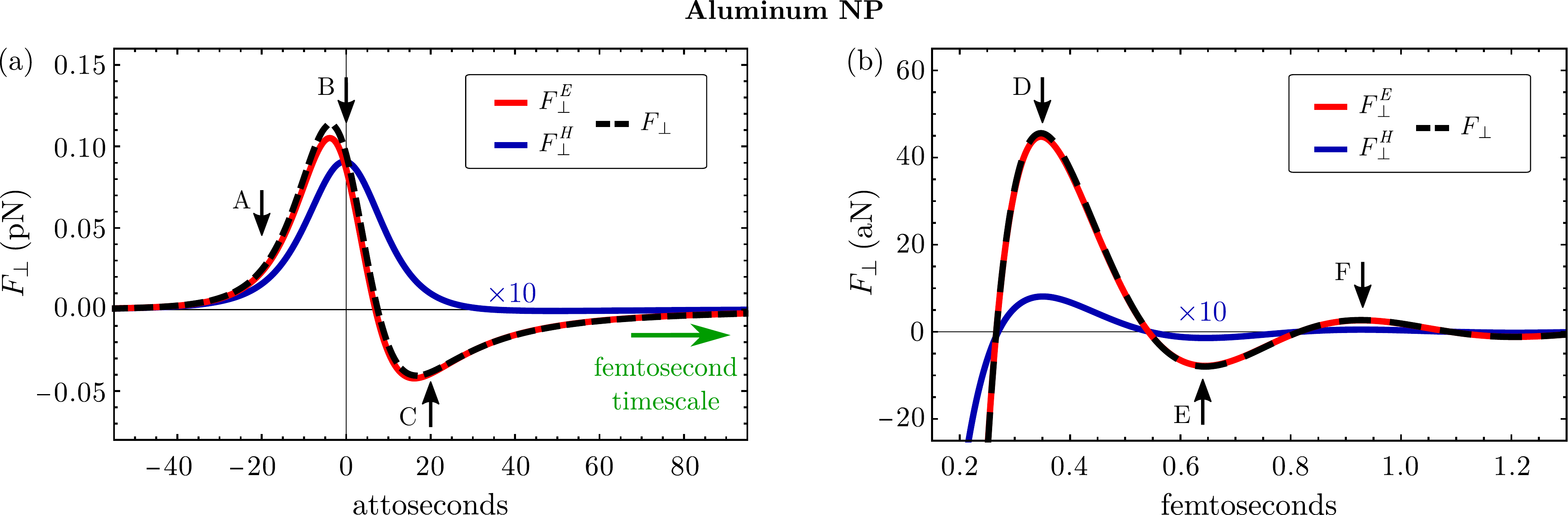} 
\caption{(a) Attosecond timescale of $F_{\bot}$ calculated for an aluminum NP with $a=1\,\text{nm}$. The swift electron is traveling with $v=0.5c$ and $b=3\,\text{nm}$. The red line represents the electric contribution to the force, the blue line represents the magnetic contribution to the force, and the black dashed line is the total force (sum of electric and magnetic contributions). Labels \textbf{A}, \textbf{B} and \textbf{C} indicate the forces on the NP at times $-20$, $0$ and $20\,\text{as}$. The horizontal green arrow points to the femtosecond  timescale. (b) Femtosecond timescale of $F_{\bot}$ for the same NP as in (a). Labels \textbf{D}, \textbf{E} and \textbf{F} in (b) indicate the forces at times $350$, $640$ and $930\,\text{as}$, respectively. The blue curves in (a) and (b) are multiplied by a factor of $10$.}
\label{Fig. Fp_Al}
\end{figure*}

\section{Interaction forces and electromagnetic radiation emitted by metallic nanoparticles}

In this Section we calculate the interparticle (swift electron and NP) forces within the dipole approximation for two different NPs: one made of aluminum and another one made of gold. Additionally, we obtain the electromagnetic radiation fields emitted by the NP due to the swift electron, and calculate the energy and linear momentum radiated by the NP.

\subsection{Force on the nanoparticle}


First, we consider the case of an aluminum NP characterized by a dielectric function within the Drude model, with a plasma frequency $\hbar \omega_p =  13.14\,\text{eV}$ and a damping parameter $\hbar \Gamma = 0.197\,\text{eV}$.\cite{Rakic} The NP has radius $a=1$\,nm and the swift electron travels with constant speed $v=0.5c$ and impact parameter $b=3\,\text{nm}$. It is worth to mention that for these parameters the aluminum NP is well described by the dipole approximation, as we will show in Section IV. The transverse or \textit{x}-component [relative to the swift electron trajectory, see Fig. \ref{Fig1}(a)] of the force $\vec{F}$ on the NP provides information of the attraction or repulsion of the NP relative to the electron beam. Thus, we focus our analysis and calculations on this component.

\begin{figure}
\centering
\includegraphics[width=.49\textwidth]{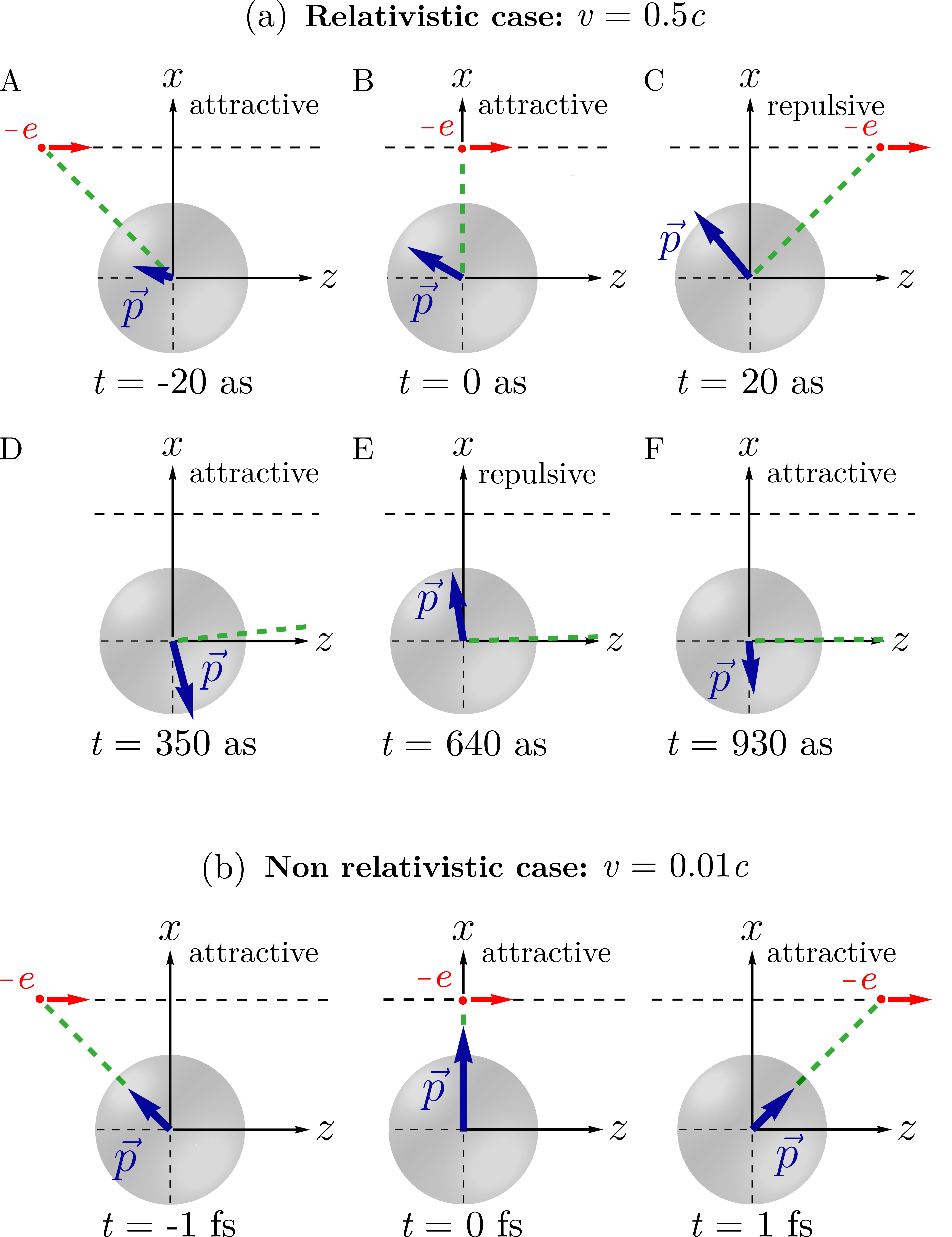} 
\caption{(a) Induced electric dipole moment (blue arrow) depicted in the plane $y=0$, for the six different times indicated in Fig. \ref{Fig. Fp_Al} by labels \textbf{A} to \textbf{F}. The swift electron (red dot) is traveling with $v=0.5c$ and $b=3\,\text{nm}$ nearby an aluminum NP with $a=1$\,nm. The green dashed line joins the red dot and the origin. (b) Analogous to (a) but now $v=0.01c$.}
\label{Fig. dip_al}
\end{figure}
%

\begin{figure*}
\includegraphics[width=\textwidth]{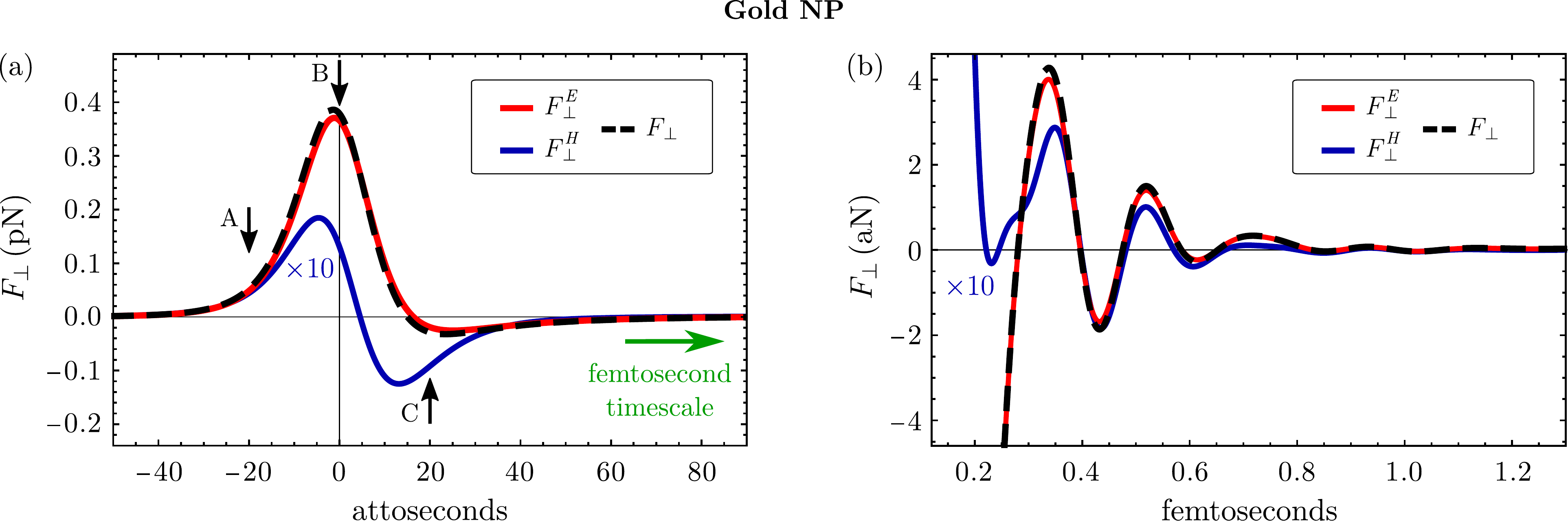}
\caption{(a) Attosecond timescale of $F_{\bot}$ calculated for a gold NP with $a=1\,\text{nm}$. The swift electron is traveling with $v=0.5c$ and $b=3\,\text{nm}$. The red line represents the electric contribution to the force, the blue line represents the magnetic contribution to the force, and the black dashed line is the total force (sum of electric and magnetic contributions). Labels \textbf{A}, \textbf{B} and \textbf{C} indicate the forces on the NP at times $-20$, $0$ and $20\,\text{as}$. The horizontal green arrow points to the femtosecond timescale. (b) Femtosecond timescale of $F_{\bot}$ for the same NP as in (a). The blue curves in (a) and (b) are multiplied by a factor of $10$.}
\label{Fig. Fp_Au}
\end{figure*}

We show in Fig. \ref{Fig. Fp_Al} the transverse component of the force ($F_{\bot}$, black dashed line) experienced by the aluminum NP.  The force is separated in its electric ($F^{\text{E}}_{\bot}$, red line) and magnetic ($F^{\text{H}}_{\bot}$, blue line) contributions [first and second terms in the \textit{rhs} of Eq. (\ref{Fp}), respectively]. From Fig. \ref{Fig. Fp_Al} one can notice two relevant time regimes for the force: (i) an attoseconds regime and (ii) a femtoseconds regime. At the attoseconds timescale the forces are in the order of piconewtons [see Fig. \ref{Fig. Fp_Al}(a)], while at the femtoseconds timescale the forces oscillate between negative and positive values and their magnitude decreases about four orders of magnitude (to tens of attonewtons) compared to the attoseconds regime [see Fig. \ref{Fig. Fp_Al}(b)].
As can be seen from Figs. \ref{Fig. Fp_Al}(a) and \ref{Fig. Fp_Al}(b), the electric contribution to the force (red line) dominates over the magnetic one (blue line, amplified tenfold). Thus, the curve of the total force (black dashed line in Fig. \ref{Fig. Fp_Al}) is superimposed to its electric contribution $F^{\text{E}}_{\bot}$. At higher speeds of the swift electron ($v>0.5c$, not shown here), the magnetic contribution to the force increases but never surpasses the electric contribution.

A positive force  ($F_{\bot} > 0$) implies that the NP is attracted towards the swift electron trajectory and, conversely, a negative force ($F_{\bot} < 0$) implies that the NP is repelled from it.
In Fig. \ref{Fig. Fp_Al}(a) we observe that at $t= -20\,\text{as}$ and at $t=0\,\text{as}$ (labels \textbf{A} and \textbf{B}, respectively) correspond to instants where the interaction between the swift electron and the NP is attractive. 
This period of time is followed by a repulsive regime (label \textbf{C}) that takes place up to $t \approx 250\,\text{as}$. After this time, the force oscillates in the femtosecond timescale between positive and negative values, leading to attractive and repulsive interactions, as shown by labels \textbf{D}, \textbf{E} and \textbf{F} in Fig. \ref{Fig. Fp_Al}(b). The repulsive interaction is caused by a delay in the response of the induced electric dipole, as we discuss in the following.
%
%
\begin{figure}
\centering
\includegraphics[width=.49\textwidth]{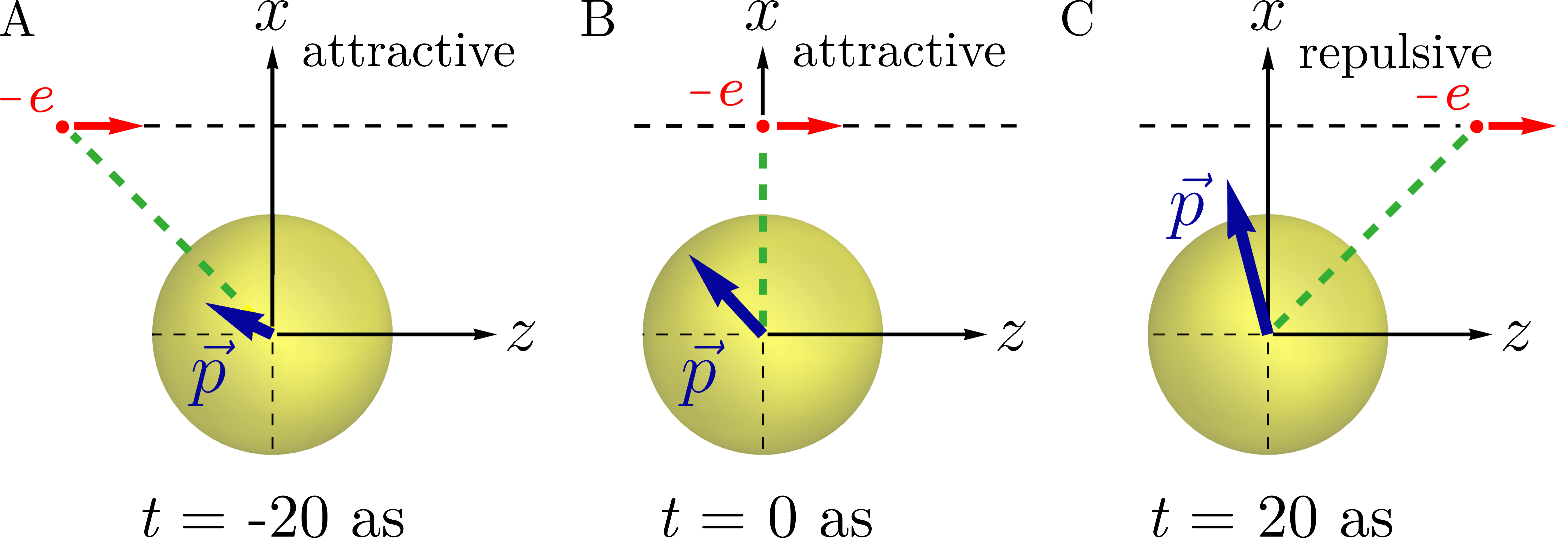} 
\caption{Induced electric dipole moment (blue arrow) depicted in the plane $y=0$, for the three different times indicated in Fig. \ref{Fig. Fp_Au} by labels \textbf{A} to \textbf{C}. The swift electron (red dot) is traveling with $v=0.5c$ and $b=3\,\text{nm}$ nearby a gold NP with $a=1$\,nm. The green dashed line joins the red dot and the origin. }
\label{Fig. dip_au}
\end{figure}
In Fig. \ref{Fig. dip_al}(a) we show six different schematics of the induced electric dipole moment $\vec{p}(t)$ within the NP (blue arrow), and the relative position of the swift electron at the six different times marked in Fig. \ref{Fig. Fp_Al}(a) and \ref{Fig. Fp_Al}(b). 
The size of the blue arrows in Fig. \ref{Fig. dip_al}  represents the magnitude of the induced dipole moment at each time ($-20$, $0$, $20$, $350$, $640$ and $930\,\text{as}$). Notice that the arrowhead represents effectively positive charge and the arrow tail represents effectively negative charge within the NP.
To visualize the delay in the response of the induced electric dipole, we trace a green dashed line joining the position of the swift electron (red dot) and the center of the NP.
At the times $t=-20\,\text{as}$ and $t=0\,\text{as}$, one can see in Fig. \ref{Fig. dip_al}(a) labels \textbf{A} and \textbf{B} that the blue arrow points behind the green dashed line. 
For these cases the effective interaction between the swift electron and the induced electric dipole is attractive\textemdash the positive induced charge of the NP is closer to the swift electron\textemdash. Conversely, for $t=20\,\text{as}$ [Fig. \ref{Fig. dip_al}(a) label \textbf{C}], the effective interaction between the swift electron and $\vec{p}$ is repulsive\textemdash at this time the negative charge of the NP is closer to the swift electron\textemdash. For later times $350$, $640$ and $930\,\text{as}$, shown in Fig. \ref{Fig. dip_al}(a) by labels \textbf{D}, \textbf{E} and \textbf{F}, respectively, the swift electron is far away and the induced dipole moment $\vec{p}$ rotates clockwise. This explains why the force experienced by the NP alternates between positive and negative in Fig. \ref{Fig. Fp_Al}(b), meaning attraction and repulsion towards the swift electron.


When the swift electron travels slow compared to $c$ (non-relativistic case), the induced dipole moment follows the position of the swift electron (red dot). That is, $\vec{p}$ points towards the red dot at all times, as we show in Fig. \ref{Fig. dip_al}(b). For this case we consider a swift electron traveling with $v=0.01c$ and $b=3\,\text{nm}$. 
We show three different times: $-1$, $0$, and $1\,\text{fs}$. One can observe in Fig. \ref{Fig. dip_al}(b) that $\vec{p}$ and the green dashed line are superimposed. For this situation, the force between the swift electron and the NP is always attractive. From this analysis we conclude that, if the swift electron travels with a speed comparable with the speed of light, a delay occurs in the response of the NP to the external electromagnetic fields, which leads to repulsive forces on the NP (see Fig. \ref{Fig. Fp_Al}). 


\begin{figure}
\centering
\includegraphics[width=0.48\textwidth]{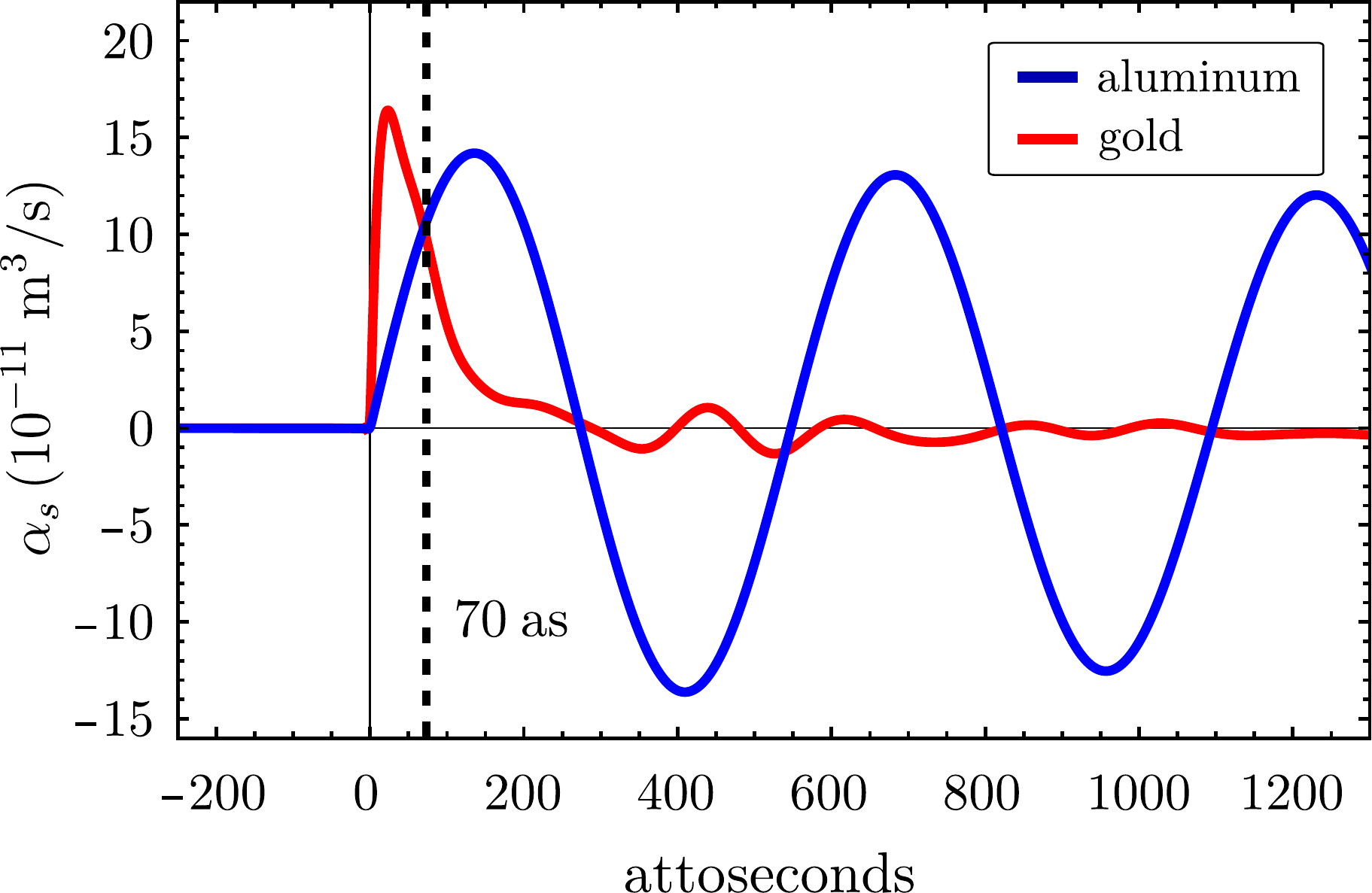} 
\caption{Quasi-static polarizability $\alpha_s (\tau)$ as a function of time, calculated for an aluminum NP with $a=1$\,nm (blue line) and for a gold NP with $a=1$\,nm (red line). The vertical black dashed line indicates $\tau=70$\,as.}
\label{Fig. Polart}
\end{figure}


Next, we consider the case of a NP made of gold with a dielectric function taken from reference\setcitestyle{numbers,square}\citep{Werner}. In Fig. \ref{Fig. Fp_Au} we show the transverse component of the force experienced by the gold NP as a function of time. The swift electron is traveling with $v=0.5c$ and $b=3\,\text{nm}$. 
Some similarities between Fig. \ref{Fig. Fp_Al} and \ref{Fig. Fp_Au} become apparent. For instance, we identify again two relevant time regimes for the force: attoseconds and femtoseconds. At the attosecond timescale the force changes from positive to negative between $-20$ to $20\,\text{as}$ [see black dashed line in Fig. \ref{Fig. Fp_Au}(a)] with a magnitude in the order of piconewtons. At the femtosecond timescale, the force decreases five orders of magnitude (attonewtons) and oscillates between positive and negative values. The amplitude of this oscillation monotonically decreases to zero.
Notice that for both timescales [Fig. \ref{Fig. Fp_Au}(a) and Fig. \ref{Fig. Fp_Au}(b)] the electric contribution to the force (red line) dominates over the magnetic one (blue line, amplified tenfold).

The repulsion (negative force) of the NP from the swift electron is caused by the delay in the response of the induced electric dipole, as we show in Fig. \ref{Fig. dip_au}.
We observe that at $-20\,\text{as}$ and $0\,\text{as}$ the effective positive induced charge of the NP (head of the blue arrow in Fig. \ref{Fig. dip_au} labels \textbf{A} and \textbf{B}) is closer to the swift electron. Thus, the force is attractive. In contrast, at $20\,\text{\,\text{as}}$ the effective induced charge of the NP is negative (tail of the blue arrow in Fig. \ref{Fig. dip_au} label \textbf{C}) which is closer to the swift electron, leading to a repulsive force. We learn from Figs. \ref{Fig. Fp_Al}-\ref{Fig. dip_au} that the force experienced by the aluminum NP has a similar behaviour, but a different magnitude, to the force experienced by the gold NP. To understand the difference in the magnitude of the forces (compare Fig. \ref{Fig. Fp_Al} and \ref{Fig. Fp_Au}), we analyze the quasi-static polarizability for both materials below.

By substituting the Drude dielectric function into Eq. (\ref{staticpol}) and performing a frequency-to-time Fourier transform of the quasi-static polarizability $\alpha_s(\omega) \rightarrow \alpha_s(\tau)$, one obtains:
\begin{equation}
\alpha_s (\tau)= 4 \pi a^3 \Theta(\tau) \dfrac{\omega_s^2}{\Omega_s} e^{-( \tau \Gamma ) /2} \sin(\Omega_s \tau),
\label{alphaAl}
\end{equation}
for the aluminum NP.\setcitestyle{numbers,super}\citep{Jackson} The $\Theta(\tau)$ in Eq. (\ref{alphaAl}) is the Heaviside step function, $\omega_s=\omega_p/\sqrt{3}$, and 
\begin{equation}
\Omega_s = \omega_s \sqrt{1-(\Gamma/2\omega_s)^2}.
\label{Omegas}
\end{equation}
In Fig. \ref{Fig. Polart} we show $\alpha_s (\tau)$ calculated for the aluminum NP (blue line). The frequency of the oscillations of $\alpha_s (\tau)$ is equal to $\Omega_s$ [given by Eq. (\ref{Omegas})]. 
It is worth noting that $\vec{p}(t)$ is obtained as a convolution of $\alpha_s (\tau)$, an oscillating function in time, and the external non-oscillating electric field [given by Eq. (\ref{Et})]. Hence, $\vec{p}$ oscillates at the same frequency $\Omega_s$. Additionally, the force on the NP [Eq. (\ref{Fp})] is proportional to $\vec{p}$ (and $\dot{\vec{p}}$\,), hence the frequency at which the force oscillates at the femtosecond regime is $\Omega_s$.

We also obtain $\alpha_s (\tau)$ for a gold NP through a numerical frequency-to-time Fourier transform, shown by the red line in Fig. \ref{Fig. Polart}. 
One recognizes that $\alpha_s (\tau)$ also oscillates but not with a single frequency as in the aluminum case. Most importantly, the magnitude of $\alpha_s$ at $\tau<70\,\text{as}$ for gold is larger than that of the aluminum (blue line).  This discrepancy in the magnitude of $\alpha_s$ explains the difference in the magnitude of the forces in Figs. \ref{Fig. Fp_Al} and \ref{Fig. Fp_Au}. Also, for the same time interval, the initial slope of $\alpha(\tau)$ for aluminum (blue curve) is smaller than the one for gold (red curve). Therefore, the induced electric dipole is delayed more in aluminum than in gold, yielding a larger repulsive-to-attractive force ratio, as can be seen by comparing Figs. \ref{Fig. Fp_Al}(a) and \ref{Fig. Fp_Au}(a).

\subsection{Force on the swift electron}

\begin{figure*}
\includegraphics[width=\textwidth]{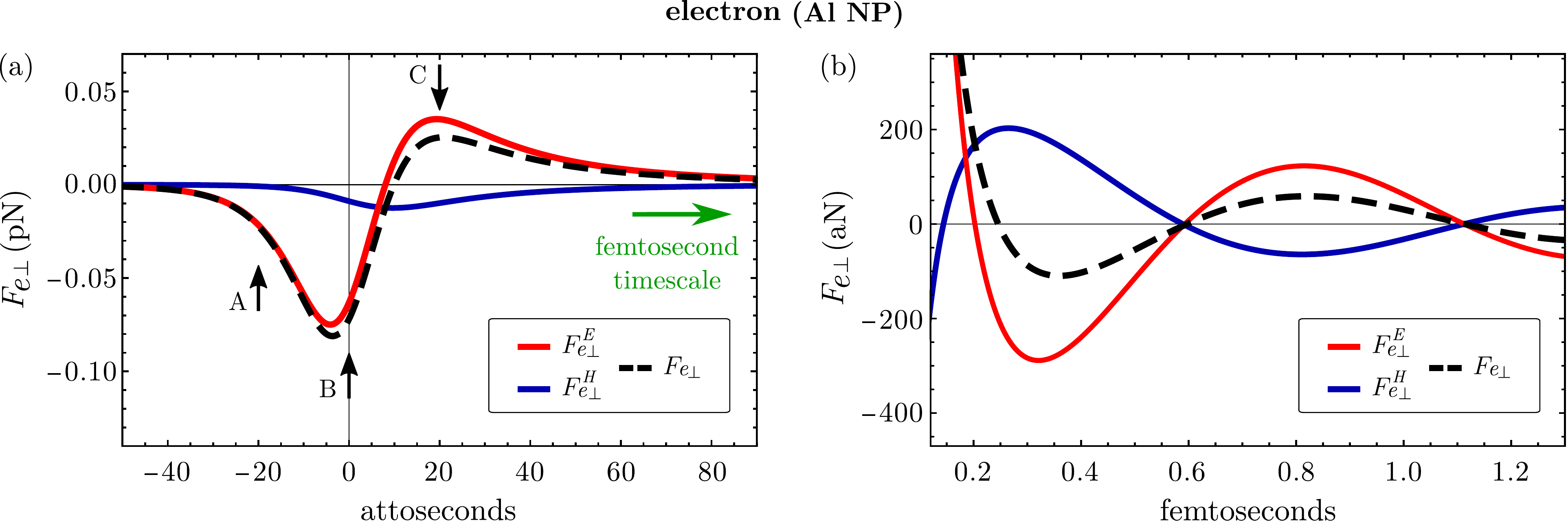}
\caption{(a) Attosecond timescale of $F_{e\bot}$ on a swift electron traveling with $v=0.5c$ and $b=3\,\text{nm}$ nearby an aluminum NP with $a=1\,\text{nm}$. The red line represents the electric contribution to the force, the blue line represents the magnetic contribution to the force, and the black dashed line is the total force (sum of electric and magnetic contributions). Labels \textbf{A}, \textbf{B} and \textbf{C} indicate the forces on the swift electron at times $-20$, $0$ and $20\,\text{as}$. The horizontal green arrow points to the femtosecond timescale. (b) Femtosecond timescale of $F_{e \bot}$ for the same swift electron as in (a).}
\label{Fig. Fe_Al}
\end{figure*}

Now, we calculate the force on the swift electron by means of Eq. (\ref{Fe}). As mentioned out before, the transverse component of the force provides information of the attraction or repulsion (deflection) of the electron beam towards the NP, hence we focus our analysis and calculations on this component.
In Fig. \ref{Fig. Fe_Al} we show the transverse component of the force on the swift electron ($F_{e \bot}$, black dashed line) when traveling at speed $v=0.5c$ with $b=3\,\text{nm}$ nearby an aluminum NP with $a=1\,\text{nm}$. We show the electric ($F^{\text{E}}_{e \bot}$, red line) and magnetic ($F^{\text{H}}_{e \bot}$, blue line) contributions to the total force [first and second terms in the \textit{rhs} of Eq. (\ref{Fe}), respectively].

When $F_{e \bot} < 0$ the swift electron is attracted towards the NP and, conversely, when $F_{e \bot} > 0$ it is repelled. We observe that the force on the swift electron (Fig. \ref{Fig. Fe_Al}) shows both attosecond and femtosecond timescales (with forces in the order of piconewtons and attonewtons, respectively). 
At the attosecond timescale, at times $t=-20\,\text{as}$ and $t=0\,\text{as}$ [labels \textbf{A} and \textbf{B} in Fig. \ref{Fig. Fe_Al}(a)], the swift electron is attracted towards the NP. Conversely, at time $t=20\,\text{as}$ [label \textbf{C} in Fig. \ref{Fig. Fe_Al}(a)] the swift electron is repelled. At the femtosecond timescale the force oscillates with frequency $\Omega_s$ between positive and negative [see Fig. \ref{Fig. Fe_Al}(b)]. This repulsive interaction is due to the delayed electric point dipole, as we discussed before from Fig. \ref{Fig. dip_al}. Additionally, from Figs. \ref{Fig. Fp_Al} and \ref{Fig. Fe_Al} we observe that the force on the swift electron presents similar behavior (but opposite direction) to the force on the NP (Fig. \ref{Fig. Fp_Al}). However, one can notice that
\begin{equation}\label{newton}
\vec{F}_\textit{e}(t) + \vec{F}(t)\neq \vec{0}.
\end{equation}
Nonetheless, the total linear momentum of the system is conserved, as we show in the following section.

\subsection{Radiation emitted by the nanoparticle and conservation of linear momentum}

\begin{figure}
\centering
\includegraphics[width=0.48\textwidth]{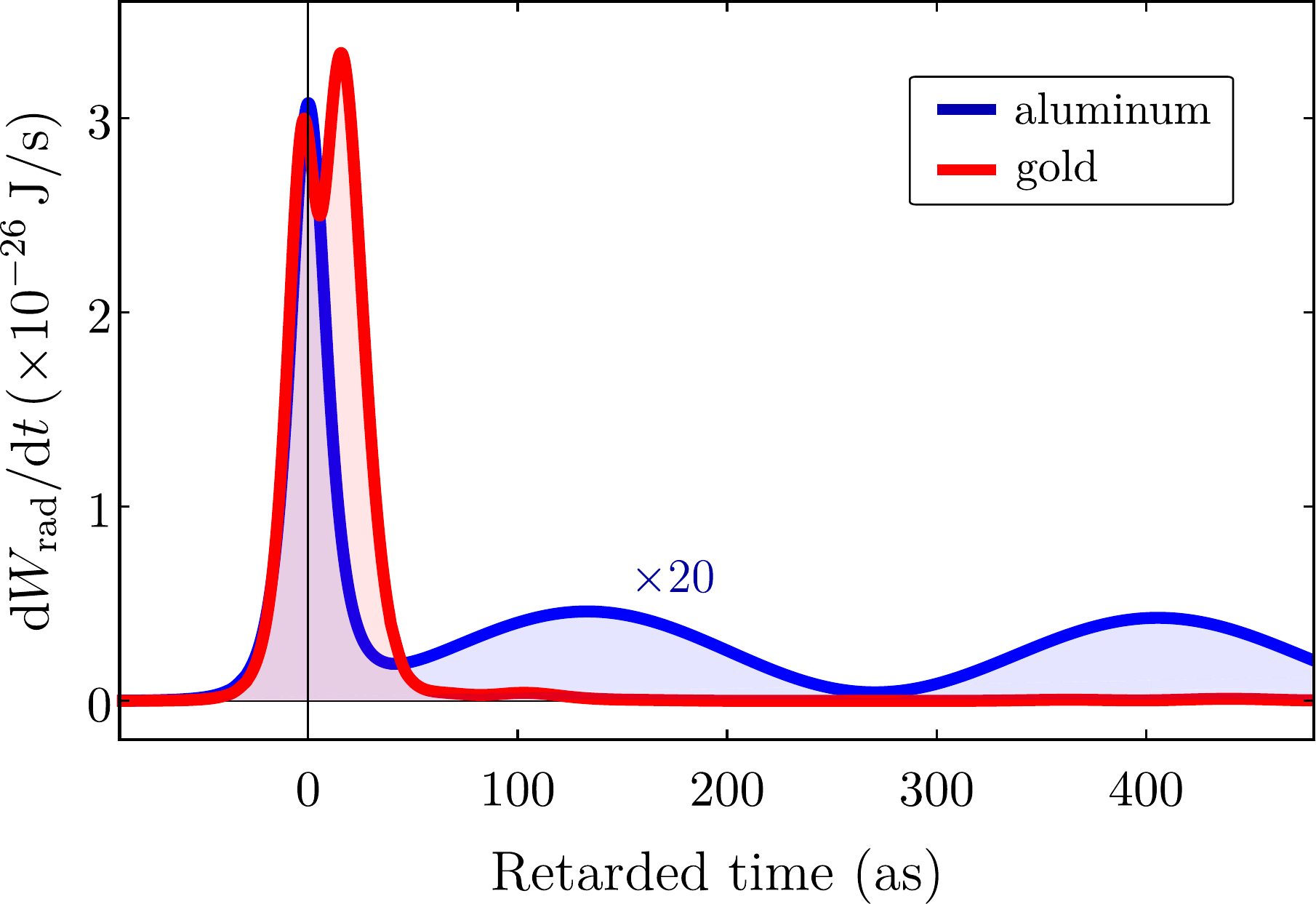} 
\caption{Rate of electromagnetic energy radiation as a function of retarded time ($t_r$), emitted by an aluminum NP with $a=1$\,nm (amplified 20 fold in blue line) and by a gold NP with $a=1$\,nm (red line). We consider a swift electron with $v=0.5c$ and $b=3$ nm. The colored area under the blue and red lines represents the energy radiated by the NP.}
\label{Fig. Urad}
\end{figure}

By considering the linear momentum that the radiation carries, we calculate the recoil by the NP when it radiates. First, we obtain the rate of energy radiation by integrating the Poynting vector $\vec{S}$ in a closed surface containing the NP:
\begin{equation}
\dfrac{\text{d} W_{\text{rad}}}{\text{d} t} = \oint \vec{S} (\vec{r}\,; t)  \cdot \text{d}\vec{a},
\label{RadEnergy}
\end{equation}
with
\begin{equation}\label{Poynting}
\vec{S}(\vec{r}\,; t)  = \vec{E}_p^{\text{rad}}(\vec{r}\,; t)  \times \vec{H}_p^{\text{rad}}(\vec{r}\,; t) ,
\end{equation}
where $\vec{E}_p^{\text{rad}}$ and $ \vec{H}_p^{\text{rad}}$ correspond to the radiation fields of the time-dependent electric point dipole $\vec{p}$ induced within the NP. The radiation fields can be extracted from Eqs. (\ref{Eps}) and (\ref{Hps}) in Appendix \ref{Apendix A}, as the terms that decay as $1/r$. Inserting $\vec{E}_p^{\text{rad}}$ and $ \vec{H}_p^{\text{rad}}$ into Eq. (\ref{Poynting}) yields
\begin{equation}\label{poynting2}
\vec{S}(\vec{r}\,; t)  = \dfrac{\mu_0}{16 \pi^2 c r^2} \left[ |\ddot{\vec{p}}\,(t_r)|^2 - 
|\hat{r} \cdot \ddot{\vec{p}} \,(t_r) |^2 \, \right] \hat{r},
\end{equation}
where $t_r = t - r/c$ is the retarded time. By calculating the closed surface integral in Eq. (\ref{RadEnergy}), one obtains: \cite{Griffiths}
\begin{equation}\label{Urad}
\dfrac{\text{d} W_{\text{rad}}}{\text{d} t} = \dfrac{\mu_0}{6 \pi c} |\ddot{\vec{p}}\,(t_r)|^2 .
\end{equation}
%
%
We show in Fig. \ref{Fig. Urad} the rate of electromagnetic energy radiated by the NP, Eq. (\ref{Urad}), as a function of the retarded time $t_r=t-r/c$, for an aluminum NP (amplified 20 fold and shown as a blue line) and for a gold NP (shown as a red line). It is clear that both NPs radiate energy (see blue and red colored area in Fig. \ref{Fig. Urad}), and that the gold NP radiates most of the energy in a short period of time (tens of attoseconds) while the radiation by the aluminum NP spreads over a longer period of time. However, from Eq. (\ref{poynting2}), we realize that $\vec{S}$ has the following symmetry:
\begin{equation}
\vec{S}(\vec{r}\,) = - \, \vec{S}(-\vec{r}\,),
\end{equation}
thus, the NP radiates the same amount of energy in one direction than in the opposite direction. From the relation between the Poynting vector and the density of electromagnetic linear momentum, $\vec{g} = \vec{S}/c^2$, we conclude that in the dipole approximation there is no radiation-reaction force on the NP.

Additionally, we calculate the total linear momentum lost by the swift electron and transferred to the NP by integrating numerically the forces [Eqs. (\ref{DPp}) and (\ref{DPe})], obtaining that
\begin{equation}\label{newton2}
\Delta \vec{P}_\textit{e} + \Delta \vec{P} = \vec{0}
\end{equation}
for both aluminum and gold NPs. Thus, the total linear momentum gained by the NP is precisely the same that the linear momentum lost by the swift electron. 

It is well known that for small NPs it is necessary to consider the size corrections to the dielectric function\citep{noguez} (due to the reduction of the mean free path of electrons inside the NP). However, we have corroborated (not shown) that they yield negligible contribution to the linear momentum transfer.

\section{Comparison of the dipole approximation with the full-retarded wave solution}\label{HA}

In this Section we obtain the total linear momentum $\Delta \vec{P}$ using the full-retarded wave solution (briefly presented below). By comparing $\Delta \vec{P}$ obtained with both, the full-retarded wave solution and the dipole approximation, we find the conditions in which the dipole approximation is valid, as we show in the following.


As we mentioned in the Introduction, the full-retarded wave solution corresponds to the exact solution of Maxwell's equations in frequency space. In this scheme, the electromagnetic fields scattered by the spherical NP are obtained as a multipolar expansion:\citep{GAbajo99}
\begin{align}
\vec{E}^{\text{scat}} (\vec{r};\omega) = \sum_{\ell = 1}^{\infty} \sum_{m=-\ell}^{m=\ell}
E_{\ell, m}^{\,r} \hat{r} + E_{\ell, m}^{\,\theta} \hat{\theta} + E_{\ell ,m}^{\,\phi} \hat{\phi}, \label{Escat} \\
\vec{H}^{\text{scat}} (\vec{r};\omega) = \sum_{\ell = 1}^{\infty} \sum_{m=-\ell}^{m=\ell}
H_{\ell, m}^{\,r} \hat{r} + H_{\ell, m}^{\,\theta} \hat{\theta} + H_{\ell, m}^{\,\phi} \hat{\phi}. \label{Hscat} 
\end{align}
We refer the reader to Appendix \ref{Apendix A}, where we show the relationship between the spherical components of Eqs. (\ref{Escat}) and (\ref{Hscat}) with the elements of the scattering-matrix $t_{\ell}^E$ and $t_{\ell}^M$ [Eqs. (\ref{ScatMatrixE}) and (\ref{ScatMatrixM})].

The total linear momentum that the swift electron transfers to the NP is given by
\begin{equation} 
\label{momentum}
\Delta\vec{P}
=
\int_{0}^{\infty} \vec{\mathcal{P}}(\omega) \, \text{d}\omega ,
\end{equation}
where
\begin{align}
\label{spectral}
\vec{\mathcal{P}}(\omega)
=&
\frac{1}{\pi}
\oint_S 
\Re \left[
\varepsilon_0\vec{E}\left(\vec{r};\omega\right)\vec{E}^{*}\left(\vec{r};\omega\right) \right. \nonumber\\ 
& - \frac{\varepsilon_0}{2}\tensor{\text{I}}\vec{E}\left(\vec{r};\omega\right)\cdot\vec{E}^{*}\left(\vec{r};
\omega\right) 
+\mu_0\vec{H}\left(\vec{r};\omega\right)\vec{H}^{*}\left(\vec{r};\omega\right) \nonumber\\ 
&\left.-
\frac{\mu_0}{2}\tensor{\text{I}}\vec{H}\left(\vec{r};\omega\right)\cdot\vec{H}^{*}\left(\vec{r};\omega\right)
\right]
\cdot \text{d}\vec{a},
\end{align}
%
with $S$ a closed surface that contains the NP and does not intersect the swift electron path. The $\Re[z]$ denotes the real part of $z$, $\mu_0$ is the vacuum permeability, $\tensor{\text{I}}$ is the identity tensor, and $\vec{E}\left(\vec{r};\omega\right)$ and $\vec{H}\left(\vec{r};\omega\right)$ are the total electromagnetic fields: the sum of those produced by the swift electron [Eqs. (\ref{ExternalE}) and (\ref{ExternalH}) in Appendix \ref{Apendix A}] and those scattered by the NP [Eqs. (\ref{Escat}) and (\ref{Hscat})]. We refer the reader to  Appendix \ref{Appendix B} for a detailed derivation of Eqs. (\ref{momentum}) and (\ref{spectral}).

In practice, the scattered fields are evaluated up to a fixed multipolar contribution $\ell_{\text{max}}$ sufficiently large to achieve numerical convergence. 
Since we are interested in a solution for small particles, we consider $\ell_{\text{max}} = 1$, and expanding $t_{\ell}^E$ and $t_{\ell}^M$ [Eqs. (\ref{tE}) and (\ref{tM})] as a power series, we obtain:\citep{Bohren}
\begin{align}
t_1^{\text{E}} =& \dfrac{2 x^3}{3} \dfrac{m^2 -1 }{m^2 + 2} +  \dfrac{2 x^5}{5} \dfrac{(m^2 -1)(m^2 -2) }{(m^2 + 2)^2} 
 + O(x^6), \label{t1ESeries}\\
t_1^{\text{M}} =& \dfrac{x^5}{45}(m^2 -1) + O(x^7),
\label{t1MSeries}
\end{align}
with $x= \omega a/ c$ the size parameter and $m^2 = \varepsilon (\omega) /\varepsilon_0$. Keeping terms up to $x^3$ in Eqs. (\ref{t1ESeries}) and (\ref{t1MSeries}), one notices that only $t_{1}^E$  contributes to the scattered fields:
\begin{equation}
t_1^{\text{E}} \approx \dfrac{k^3}{6 \pi} \alpha_s(\omega), \label{tEs}
\end{equation}
where $\alpha_s(\omega)$ is the quasi-static polarizability given by Eq. (\ref{staticpol}). The other terms in Eqs. (\ref{t1ESeries}) and (\ref{t1MSeries}) are usually known as radiation corrections. 
We refer to the case in which $t_{\ell=1}^E$ is determined by Eq. (\ref{tEs}) as the small particle approximation (SPA).
In the following, we compare the transverse component of total linear momentum ($\Delta P_{\bot}$, transferred by the swift electron to the NP) obtained with the dipole approximation,  SPA, and the full-retarded wave solution with $\ell_{\text{max}} =1$ and $\ell_{\text{max}}=30$.


\begin{figure}
\includegraphics[width=.48\textwidth]{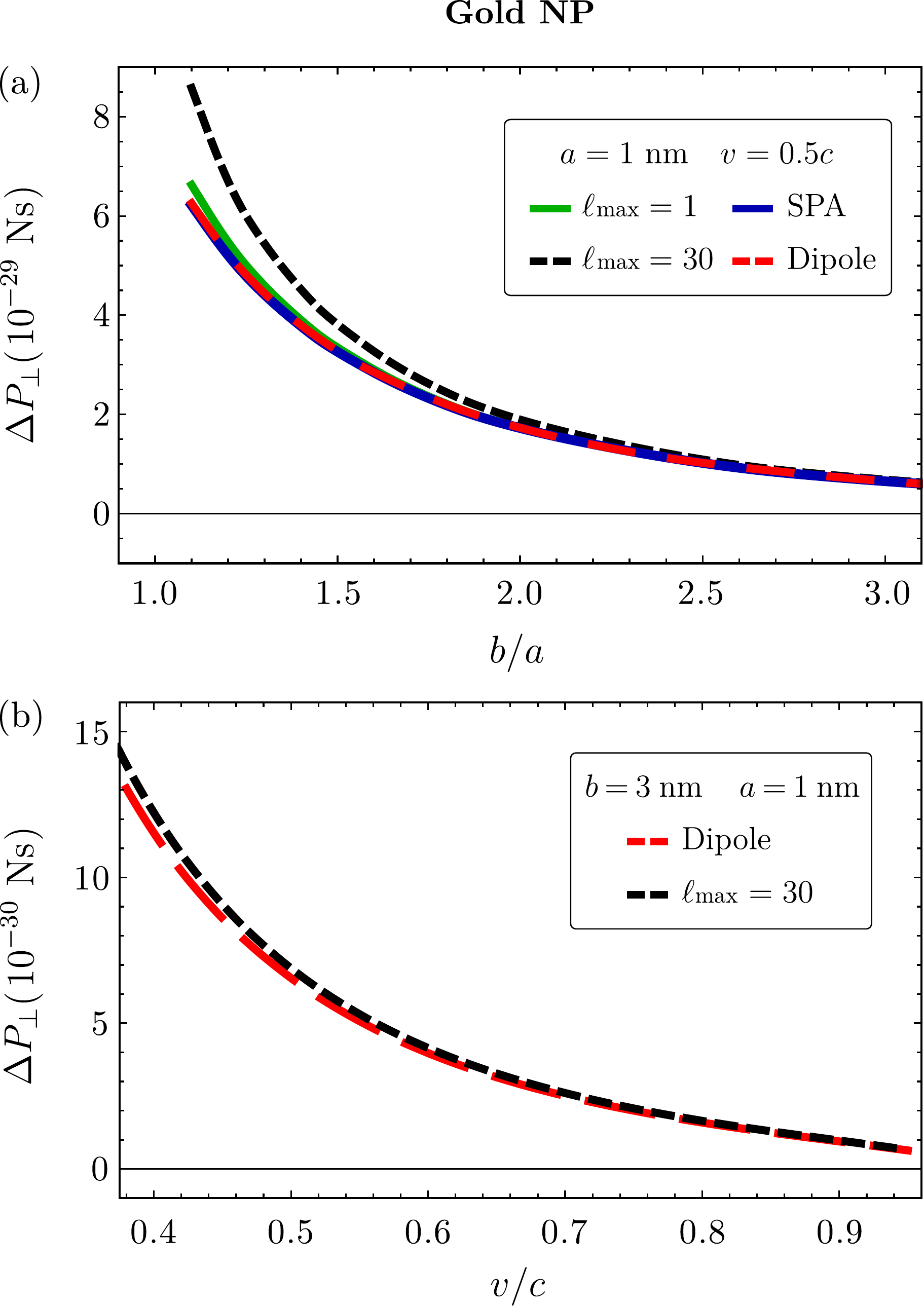} 
\caption{Transverse component of the total linear momentum transferred,  $\Delta P_{\bot}$, by a swift electron to a gold NP with $a=1\,\text{nm}$. The green line and the black dashed line correspond to $\Delta P_{\bot}$ obtained with the full-retarded wave solution with $\ell_\text{max}=1$ and $\ell_\text{max}=30$, respectively. The blue line corresponds to $\Delta P_{\bot}$ obtained with SPA, and the red dashed line to $\Delta P_{\bot}$ calculated with the dipole approximation.  
(a) Shows $\Delta P_{\bot}$ as a function of the ratio $b/a$ for a swift electron traveling with  $v=0.5c$. 
(b) Shows $\Delta P_{\bot}$ as a function of the ratio $v/c$ for a fixed impact parameter $b=3\,\text{nm}$.}
\label{Fig.DP}
\end{figure}

In Fig. \ref{Fig.DP}(a) we show $\Delta P_{\bot}$ as a function of $b/a$, assuming a swift electron traveling with constant speed $v=0.5c$, in the vicinity of a small gold NP of radius $a = 1\,\text{nm}$. 
We compare $\Delta P_{\bot}$ calculated using our time-dependent approach [the dipole approximation, Eq. (\ref{DPp})], red dashed line with: (i) SPA [Eq. (\ref{tEs})], blue line; (ii) the full-retarded wave solution with $\ell_{max} =1$ [Eq. (\ref{momentum})], green line; and (iii) the full-retarded wave solution with $\ell_{\text{max}}=30$ [Eq. (\ref{momentum})], black dashed line.
We notice two general features shared by the four calculations: (i) $\Delta P_{\bot}$ is always positive and (ii) $\Delta P_{\bot}$ decreases for larger values of $b/a$ [see Fig. \ref{Fig.DP}(a)]. From these results, we conclude that the NP is attracted towards the electron beam and the closer the swift electron is to the NP the larger the linear momentum transferred. When the impact parameter $b$ is comparable to the NP radius $a$, the higher-order multipolar modes ($\ell>1$) contribute considerably to $\Delta P_{\bot}$ (compare black dashed line with the green one). As the impact parameter to radius ratio increases, the four $\Delta P_{\bot}$ calculations converge asymptotically to the same value. 
We recognize that $\Delta P_{\bot}$ obtained with the dipole approximation superimposes $\Delta P_{\bot}$ obtained with the SPA for all $b/a$ ratios [compare red dashed line with the blue line in Fig. \ref{Fig.DP}(a)]. These findings imply that the dipole approximation is the time counterpart of the small particle approximation (SPA). The results obtained in Fig. \ref{Fig.DP}(a) allow us to define a regime where high order multipolar contribution ($\ell>1$), as well as radiation corrections, are negligible. We refer to this as the dipolar regime.

To corroborate that the dipolar regime remains valid for different swift electron speeds, we chose $b/a = 3$ and calculate $\Delta P_{\bot}$ as a function of $v/c$ for the dipole approximation and the full-retarded wave solution ($\ell_{max}=30$). 
The results are shown in Fig. \ref{Fig.DP}(b), where we notice that $\Delta P_{\bot}$ for the dipole approximation (red dashed line) follows closely the full-retarded wave solution (black dashed line) for a broad range of $v/c$ values.


\section{Conclusions}

We studied the interaction in time between a swift electron and a small metallic nanoparticle by assuming the latter as an electric point dipole (dipole approximation). Within this approach, we calculated the forces between the swift electron and the nanoparticle as a function of time and we identified two different timescales: an attosecond timescale with forces in the order of piconewtons, and a femtosecond timescale with forces in the order of attonewtons. 
Our analysis shows that the delayed response of the induced electric point dipole leads to repulsive forces.
When the electron travels slow compared to the speed of light, the delay is negligible and therefore the forces are always attractive. 

In addition, we studied the electromagnetic radiation emitted by the nanoparticle. Our results show that, within the dipole approximation, there is no radiation-reaction force on the nanoparticle due to the symmetry of the Poynting vector carried by the scattered electromagnetic fields. We corroborated that the total linear momentum gained by the nanoparticle is precisely the one lost by the swift electron, hence the total linear momentum of the system is conserved. 

To establish the validity of the dipole approximation, we compared the total linear momentum obtained from: (i) the dipole approximation and (ii) the full-retarded wave solution. We show that for large impact parameters-to-radius ratio, the dipole approximation is in good agreement with the full-retarded wave solution. 
Besides the simplicity of the dipole approximation, our findings shows that this simple model exhibits important details of the interaction between small nanoparticles and swift electrons in time domain, contributing to the understanding of fundamental interactions.

\begin{acknowledgments}
This work was supported by UNAM-PAPIIT project DGAPA IN114919. J. C-F. and J. \'A. C-R acknowledge PhD scholarships from Consejo Nacional de Ciencia y Tecnolog\'ia (CONACyT), Mexico. C. M-E acknowledges master scholarship from Consejo Nacional de Ciencia y Tecnolog\'ia (CONACyT), Mexico. 
\end{acknowledgments}

\appendix

\begin{widetext}
\section{Electromagnetic fields scattered by the nanoparticle and produced by the swift electron} \label{Apendix A}

In this appendix we present the electromagnetic fields scattered by the NP and produced by the swift electron.

When the NP is modelled as an electric point dipole (dipole approximation), the scattered electromagnetic fields are determined by the following expressions:\citep{Griffiths}
\begin{align}
\vec{E}_p (\vec{r}\,;t) =& \dfrac{1}{4 \pi \varepsilon_0 r^3} \left\{  \left[ 3 \hat{r}\hat{r} - \tensor{\text{I}} \, 
\right] \cdot \left[ \vec{p}\,(t_r) + (r/c) \, \dot{\vec{p}}\, (t_r) \, \right] + (r/c)^2 \left[
\ddot{\vec{p}}\,(t_r) \x \hat{r} \right] \x \hat{r} \right\}, \label{Eps} \\
\vec{H}_p (\vec{r}\,;t) =& - \dfrac{1}{4 \pi r^2} \, \hat{r} \x \left[ 
\dot{\vec{p}}\,(t_r) + (r/c)\,\ddot{\vec{p}}\, (t_r)\,\right], \label{Hps}
\end{align}
where $\tensor{\text{I}}$ is the identity tensor, $\vec{p}$ is the electric dipole moment, and $\dot{\vec{p}}$, $\ddot{\vec{p}}$ are time derivatives of $\vec{p}$. Notice that in Eqs. (\ref{Eps}) and (\ref{Hps}), $\vec{p}$, $\dot{\vec{p}}$, and $\ddot{\vec{p}}$ are evaluated at the retarded time $t_r = t - r/c$.

The time-dependent electromagnetic fields produced by the swift electron are:\citep{Jackson}
\begin{align} 
\vec{E}^{\text{ext}}(x,y,z;t) =& \dfrac{-e}{4{\pi}{\varepsilon}_0 } \frac{ \gamma \left[ (x-b)\hat{x}  + y \hat{y} + (z-vt) \hat{z}\right]}
{[{(x-b)^2+y^2+\gamma}^2(z-vt)^2]^{3/2}}, \label{Et} \\
\nonumber \\
\vec{H}^{\text{ext}}(x,y,z;t) =& \dfrac{-e}{4\pi}\frac{\gamma \, v \left[(x-b) \hat{y} - y\hat{x} \right]} 
{[{(x-b)^2+y^2+\gamma}^2(z-vt)^2]^{3/2}}, \label{Ht}
\end{align}
with $\gamma=(1- \beta^2)^{-1/2}$ the Lorentz factor and $\beta=v/c$. By a time-to-frequency Fourier transform of Eqs. (\ref{Et}) and (\ref{Ht}), one finds that the electromagnetic fields produced by the swift electron (in cgs units) can be expressed as follows:\citep{MacielWave}
\begin{align}
\vec{E}^{\text{ext}} \left( \vec{r} \, ; \omega \right) =& \, -\dfrac{2 e \omega}{v^2 \gamma} e^{i \, \omega (z/v)} \left\{ \frac{ \text{sign}(\omega)}{R} 
 \, K_1  \left( \frac{ |\omega| R}{v \gamma}\right) \left[ \,  (x-b) \hat{x} +  \, y \, \hat{y} \right]  -\dfrac{i}{\gamma} \,  K_0  
 \left( \frac{ |\omega| R }{v \gamma} \right) \hat{z} \right\},  \label{ExternalE}  \\
\nonumber \\
\vec{H}^{\text{ext}} (\vec{r} \, ; \omega) =& \, \dfrac{2 e \beta }{R v^2 \gamma } |\omega| e^{i \, \omega (z/v)} 
\, K_1  \left( \frac{ |\omega| R}{v \gamma}\right) \left[ \,  y \hat{x} -  \,(x-b) \, \hat{y} \right] , \label{ExternalH}
\end{align}
with $K_1(x)$ and $K_2(x)$ the modified Bessel functions of the second kind of order 1 and 2, respectively, and $R=\sqrt{(x-b)^2 + y^2}$.

The complete electromagnetic field scattered by the NP (beyond the dipole approximation) can be obtained by solving the full-retarded Maxwell's equations in frequency space. The details for the derivation of the scattered electromagnetic fields are given in reference\setcitestyle{numbers,square}\citep{GAbajo99}. Here we show the spherical multipolar components (outside the NP) of the Eqs. (\ref{Escat}) and (\ref{Hscat}) in cgs units:
\begin{align} 
{E}_{\ell, m}^{\,r} =& \, e^{i m \phi}  D^{\text{scat}}_{\ell, m}  \ell(\ell+1) P_{\ell }^m 
(\cos \theta) \dfrac{h_{\ell }^{(+)}(k_0 r)}{k_0 r}, \label{ScatEr}\\
{E}_{\ell, m}^{\, \theta} =&  -e^{i m \phi} C^{\text{scat}}_{\ell, m} \frac{m}{ \sin \theta} h_{\ell }^{(+)}(k_0 r) P_{\ell }^{m} 
(\cos \theta) - e^{i m \phi} D^{\text{scat}}_{\ell, m} \left[  (\ell+1) \dfrac{\cos \theta}{\sin \theta} P_{\ell }^{m} (\cos \theta) 
\right.  \nonumber \\
& \left. -\frac{ (\ell-m+1) }{\sin \theta} P_{\ell +1}^{m} (\cos \theta) \right] \left[ (\ell+1) \dfrac{h_{\ell }^{(+)}
(k_0 r)}{k_0 r} - h_{\ell +1}^{(+)}(k_0r) \right], \label{ScatEtheta} \\
{E}_{\ell, m}^{\,\phi} =& \, ie^{i m \phi} C^{\text{scat}}_{\ell, m} h_{\ell }^{(+)}(k_0 r) \left[ (\ell+1) \frac{\cos \theta}
{\sin \theta} P_{\ell }^{m} (\cos \theta) - \frac{(\ell-m+1)}{\sin \theta} P_{\ell +1}^{m}(\cos \theta)  \right] \nonumber \\
&+ ie^{i m \phi} D^{\text{scat}}_{\ell, m} \frac{m}{\sin \theta}  P_{\ell }^{\,m} (\cos \theta) \left[ (\ell+1) \frac{h_{\ell }^{(+)}
(k_0 r)}{k_0 r} - h_{\ell +1}^{(+)} (k_0 r) \right], \label{ScatEphi}
\end{align}
and the components of the H field are:
\begin{align}
{H}_{\ell, m}^{\,r} =& \, e^{i m \phi}  C^{\text{scat}}_{\ell, m}  \ell(\ell+1) P_{\ell }^m 
(\cos \theta) \dfrac{h_{\ell }^{(+)}(k_0 r)}{k_0 r}, \label{ScatHr}\\
{H}_{\ell, m}^{ \,\theta} =& \, e^{i m \phi} D^{\text{scat}}_{\ell, m} \frac{m}{ \sin \theta} h_{\ell }^{(+)}(k_0 r) 
P_{\ell }^{m} (\cos \theta) - e^{i m \phi} C^{\text{scat}}_{\ell, m} \left[  (\ell+1) \dfrac{\cos \theta}{\sin \theta} 
P_{\ell }^{m} (\cos \theta)  \right. \nonumber  \\
& \left. -\frac{ (\ell-m+1) }{\sin \theta} P_{\ell +1}^{m} (\cos \theta) \right] \left[ (\ell+1)\dfrac{h_{\ell }^{(+)}
(k_0 r)}{k_0 r} - h_{\ell +1}^{(+)}(k_0r) \right], \label{ScatHtheta}\\
{H}_{\ell, m}^{\, \phi} =& \, ie^{i m \phi} C^{\text{scat}}_{\ell ,m} \frac{m}{\sin \theta}  P_{\ell }^{\,m} (\cos \theta) 
\left[ (\ell+1) \frac{h_{\ell }^{(+)}(k_0 r)}{k_0 r} - h_{\ell +1}^{(+)} (k_0 r)  \right]  \nonumber \\ 
&- ie^{i m \phi} D^{\text{scat}}_{\ell, m} h_{\ell }^{(+)}(k_0 r) \left[ (\ell+1) \frac{\cos \theta} {\sin \theta} 
P_{\ell }^{m} (\cos \theta) - \frac{(\ell-m+1)}{\sin \theta} P_{\ell +1}^{m} (\cos \theta)  \right], \label{ScatHphi}  
\end{align}
where $k_0$ is the wave number in vacuum, $h_{\ell }^{(+)}(z)=ih_{\ell }^{(1)}(z)$ with $h_{\ell }^{(1)}(z)$ the spherical Hankel function of the first kind, and $P_{\ell }^{\,m}(x)$ are the associated Legendre functions. The spherical coordinate system ($r$,$\theta$,$\phi$) is determined by the Spherical-to-Cartesian transformation: $x=r \sin \theta \cos\phi$, $y=r \sin \theta \sin\phi$, $z=r\cos\theta$, with $(x,y,z)$ the Cartesian coordinate system shown in Fig \ref{Fig1}. 
The coefficients $C^{\text{scat}}_{\ell m}$ and $D^{\text{scat}}_{\ell m}$ in Eqs. (\ref{ScatEr}) to (\ref{ScatHphi}) are determined by:
\begin{align}
C^{\text{scat}}_{\ell, m} &= (i)^{\ell} \sqrt{  \frac{ 2{\ell}+1}{4 \pi} \frac{({\ell}-m)!}{({\ell}+m)!}} \,  \, \psi_{\ell,m}^{\text{M},\text{scat}}, \label{Cscatt} \\
D^{\text{scat}}_{\ell, m} &= (i)^{\ell} \sqrt{  \frac{ 2{\ell}+1}{4 \pi} \frac{({\ell}-m)!}{({\ell}+m)!}} \,  \, \psi_{\ell,m}^{\text{E},\text{scat}}. \label{Dscatt}
\end{align}
The relationship between the coefficients of the external, $\psi_{\ell,m}^{\text{E,M},\text{ext}}$, and the scattered, $\psi_{\ell,m}^{\text{E,M},\text{scat}}$, scalar functions is given by the scattering-matrix:
\begin{align}
\psi_{\ell,m}^{\text{E},\text{scat}} &= t_{\ell }^{\text{E}}  \psi_{\ell,m}^{\text{E},\text{ext}}, \label{ScatMatrixE}\\
\psi_{\ell,m}^{\text{M},\text{scat}} &= t_{\ell }^{\text{M}}  \psi_{\ell,m}^{\text{M},\text{ext}}.  \label{ScatMatrixM}
\end{align}
The full expressions for $\psi_{\ell,m}^{\text{E,M},\text{ext}}$ and $\psi_{\ell,m}^{\text{E,M},\text{scat}}$, are given in references \citep{GAbajo99,GAbajo99-2}.
The elements of the scattering-matrix are independent of $m$ due to the spherical symmetry, and are given by:
\begin{align}
t_{\ell }^{\text{E}}  &=  \frac{ - j_{\ell }(x_0) \left[ x_i j_{\ell }(x_i) \right]' +  \varepsilon_i \, j_{\ell }(x_i) 
\left[ x_0 j_{\ell }(x_0) \right]' }{h_{\ell }^{(+)}(x_0) \left[ x_i j_{\ell }(x_i) \right]' - \varepsilon_i \, 
j_{\ell }(x_i) \left[ x_0 h_{\ell }^{(+)}(x_0) \right]'} , \label{tE}\\
t_{\ell }^{\text{M}}  &=  \frac{ - x_i j_{\ell }(x_0)  j_{\ell }'(x_i)  + x_0 j_{\ell }'(x_0)  j_{\ell }(x_i) }{ x_i h_{\ell }^{(+)}(x_0) 
j_{\ell }'(x_i)  - x_0 h_{\ell }^{(+)'}(x_0) j_{\ell }(x_i) },\label{tM}
\end{align}
%
%
where $ j_{\ell }(z)$ is the spherical Bessel function, $x_i = \sqrt{\varepsilon_i} \omega a/c$, and the subscripts $i$ and $0$ indicate the dielectric function inside and outside the NP, respectively. The prime denotes differentiation with respect to $x_0$ or $x_i$.

Notice that from Eqs. (\ref{ScatEr}) to (\ref{ScatMatrixM}) one obtains the relationship between the elements of the scattering matrix and the components of the scattered electromagnetic fields (as mentioned in Section IV).
\end{widetext}

\section{Total linear momentum transfer within the full-retarded wave solution approach} \label{Appendix B}
\comentar{Total linear momentum transferred by the swift electron to the nanoparticle within the full-retarded wave solution scheme}

The total linear momentum $\Delta\vec{P}$ that is transferred by the swift electron to the NP can be calculated from the linear momentum conservation law:\citep{Jackson}
%
\begin{equation} \label{conservation}
\frac{\text{d}}{\text{d}t}\left[\vec{P}_{\text{mech}}\left(t\right)+\vec{P}_{\text{EM}}\left(t\right) \right]
=\oint_S \tensor{\text{T}}\left(\vec{r};t\right) \cdot \text{d}\vec{a}\, , 
\end{equation}
where $\vec{P}_{\text{mech}}\left(t\right)$ is the mechanical linear momentum, $\vec{P}_{\text{EM}}\left(t\right)$ is the electromagnetic linear momentum, $\tensor{\text{T}}\left(\vec{r};t\right)$ is Maxwell's stress tensor, and $S$ is a surface enclosing the NP.
The electromagnetic linear momentum is
%
\begin{equation} \label{EMmomentum}
\vec{P}_{\text{EM}}\left(t\right) 
=\varepsilon_0 \mu_0 \, \int_V
 \vec{E}\left(\vec{r};t\right) \times \vec{H}\left(\vec{r};t\right) \text{d}^3r\, ,
\end{equation}
where $\vec{E}\left(\vec{r};t\right)$ is the total electric field, $\vec{H} 
\left(\vec{r};t\right)$ is the total H field, and $V$ is the volume enclosed by $S$. The Maxwell's stress tensor is given by the following expression:
\begin{align} \label{maxtensor}
\tensor{\text{T}}\left(\vec{r};t\right)
=& \,
\varepsilon_0\vec{E}\left(\vec{r};t\right)\vec{E}\left(\vec{r};t\right)
-
\frac{\varepsilon_0}{2}\tensor{\text{I}}\vec{E}\left(\vec{r};t\right)\cdot\vec{E}\left(\vec{r};t\right)\noindent\nonumber\\ 
&+
\mu_0\vec{H}\left(\vec{r};t\right)\vec{H}\left(\vec{r};t\right)
-
\frac{\mu_0}{2}\tensor{\text{I}}\vec{H}\left(\vec{r};t\right)\cdot\vec{H}\left(\vec{r};t\right). 
\end{align}
%


From Eq. (\ref{conservation}), one obtains the total linear momentum transferred ($\Delta\vec{P}$) by the swift electron to the NP as

\begin{align} \label{redconservation}
\Delta\vec{P}=
\int_{-\infty}^{\infty}
\frac{\text{d}}{\text{dt}}\vec{P}_{\text{mech}}\left(t\right) \text{d}t
=&
\int_{-\infty}^{\infty} \oint_S
\tensor{\text{T}}\left(\vec{r};t\right) \cdot \text{d}\vec{a}\, \text{d}t \nonumber \\
& - \vec{P}_{\text{EM}}(t) \bigg|_{t=-\infty}^{t=+\infty}.
\end{align}
%
For our calculations $S$ is a sphere that does not intersect the electron trajectory.
It is important to notice that, when the electron is far from the NP at $t=-\infty$, the total electromagnetic fields inside $V$ are zero. Hence, the electromagnetic linear momentum satisfies $\vec{P}_{\text{EM}}\left(t=-\infty\right) = \vec{0}$. In a similar manner, at $t=+\infty$ the swift electron is far from the NP and the external electromagnetic fields inside $V$ are zero. Furthermore, the scattered electromagnetic fields inside $V$ are zero due to dissipative processes. Thus, the total electromagnetic fields inside $V$ at $t=+\infty$ are again zero, and $\vec{P}_{\text{EM}}\left(t=+\infty\right) =\vec{0}$. Therefore, $\Delta\vec{P}$ is given by
\begin{align} \label{redconservation2}
\Delta\vec{P}=
\int_{-\infty}^{\infty}
\frac{\text{d}}{\text{dt}}\vec{P}_{\text{mech}}\left(t\right) \text{d}t
=&
\int_{-\infty}^{\infty} \oint_S
\tensor{\text{T}}\left(\vec{r};t\right) \cdot \text{d}\vec{a}\, \text{d}t.
\end{align}
%
By a time-to-frequency Fourier transform $\vec{E}(\vec{r};t)\rightarrow \vec{E}(\vec{r};\omega)$ and $\vec{H}(\vec{r};t)\rightarrow \vec{H}(\vec{r};\omega)$ in Eq. (\ref{maxtensor}), one further obtains\citep{PRB2010} from Eq. (\ref{redconservation2}) that the total linear momentum can be expressed as 
%
\begin{align} \label{momtransf2}
\Delta\vec{P}=&\frac{1}{2\pi}\int_{-\infty}^{\infty}\oint_S\bigg[
\varepsilon_0\vec{E}\left(\vec{r};\omega\right)\vec{E}^{*}\left(\vec{r};\omega\right) \nonumber\\ 
& - \frac{\varepsilon_0}{2}\tensor{\text{I}}\vec{E}\left(\vec{r};\omega\right)\cdot\vec{E}^{*}\left(\vec{r};\omega\right) 
+\mu_0\vec{H}\left(\vec{r};\omega\right)\vec{H}^{*}\left(\vec{r};\omega\right) \nonumber\\ 
&-
\frac{\mu_0}{2}\tensor{\text{I}}\vec{H}\left(\vec{r};\omega\right)\cdot\vec{H}^{*}\left(\vec{r};\omega\right)
\bigg]
\cdot \text{d}\vec{a}\,
\text{d}\omega .
\end{align}

The total electromagnetic fields as functions of time [$\vec{E}(\vec{r}\,;t)$ and $\vec{H}(\vec{r}\,;t)$] are real functions. This implies that
\begin{equation}\label{integral}
\int_{-\infty}^{\infty} \!\!\! \vec{E} ( \vec{r} ; \omega) \vec{E}^* ( \vec{r} ; 
\omega) \, d\omega = 2 \! \int_{0}^{\infty} \!\!\! \Re \left[ \vec{E} ( \vec{r} ; \omega) \vec{E}^* ( 
\vec{r} ; \omega) \right] \, d\omega
\end{equation}
and
\begin{equation}\label{integral2}
\int_{-\infty}^{\infty} \!\!\! \vec{H} ( \vec{r} ; \omega) \vec{H}^* ( \vec{r} ; 
\omega) \, d\omega = 2 \! \int_{0}^{\infty} \!\!\! \Re \left[ \vec{H} ( \vec{r} ; \omega) \vec{H}^* ( 
\vec{r} ; \omega) \right] \, d\omega
\end{equation}
due to $\vec{E}^* (\vec{r};\omega) = \vec{E} (\vec{r};- \omega)$ and  $\vec{H}^* (\vec{r};\omega) = \vec{H} (\vec{r};- \omega)$.
Therefore, from Eqs. (\ref{momtransf2}),  (\ref{integral}), and (\ref{integral2}), the total linear momentum transferred by the swift electron to the NP can be expressed as we show in Eqs. (\ref{momentum}) and (\ref{spectral}) in the main text.


%

\end{document}